\documentclass[preprint,aps,prl,showpacs]{revtex4}

\usepackage[latin1]{inputenc} 
\usepackage{amsmath}
\usepackage{amsfonts}
\usepackage{amssymb}
\usepackage{graphicx}
\usepackage{enumerate}

\newcommand{\be}{\begin{equation}}
\newcommand{\ee}{\end{equation}}

\begin{document}

\title{Clusters or networks of economies? \\ A macroeconomy study through GDP $fluctuation$ correlations}

\author{ M. Ausloos$^{\dagger}$ and R. Lambiotte$^{\dagger\dagger}$}

\email[$^{\dagger}$]{marcel.ausloos@ulg.ac.be}

\email[$^{\dagger\dagger}$]{renaud.lambiotte@ulg.ac.be}

\affiliation{  Institute of  Physics, B5, University
of Li$\grave e$ge, B-4000 Li$\grave e$ge, Euroland }

\date{\today}

\begin{abstract} 
We follow up on the study of correlations between GDP's of rich countries, as studied in refs. [1-3]. We analyze web-downloaded data on
GDP that we use as individual wealth signatures of the country economical state. We calculate the yearly fluctuations of the GDP. We look for forward and backward correlations between such fluctuations. The system is represented by an evolving  network, nodes being the GDP fluctuations (or countries) at different times. 

In order to extract structures from the network, we focus on filtering the time delayed correlations by removing the least correlated links. This percolation idea-based method reveals the emergence of connections, that are visualized by a branching representation. Note that the network is made of weighted and directed links when taking into account a delay time. Such a measure of collective habits does not fit the usual expectations defined by politicians or economists.  
\end{abstract}


\maketitle

\section{Introduction}

Some time ago,  the question of economy globalization was studied along the lines of macro-econophysics research \cite{1,2,2b}. Other considerations along related lines can be found in these proceedings in contributions by Miskiewicz \cite{3}  and Gligor \cite{4}. The question pertains to well known observations: there are political, cultural, scientific, economic cooperations all over the world, even though there are even anti-globalization organizations.  

In our case, we consider that a $globalization process$ in economy should be understood as an increase of similarities within (macroeconomy) development patterns. In so doing we have proposed to study whether the pattern economy of countries was well represented by the $distance$ between  fluctuations  of  their country Normalized Gross Domestic Products (GDP). The notion of distance is somewhat arbitrary, as has been already discussed in [1],  but is left for further investigation elsewhere.   We have chosen to present here a study based on the statistical distance, in Sec. II, 
 for the 23 most developed countries for which annual GDP data is available in the interval  $[1950 - 2003]$. 
 
 One might argue \cite{fagiolo} that some data detrending is necessary to take into account cycle like effects \cite{cycleLong83,cycleBasu99}.  In previous studies \cite{1} it has been observed that there is much noise is the data. We have shown that some average should be taken over large time windows to remove such a noise. We will use an average time of $T=4$ years. One usual $political$ question is whether one has to $follow$ another country policy in order to improve one's country economy. The question of followers and leaders can be tackle by considering time delayed correlations. In this first attempt we will only consider a one year time lag $\tau $, being aware that other time lags should be involved. Note that   positive or negative ones should be considered. Nevertheless it is worthwhile to check whether such an approach is  already meaningful for $\tau= 1$ and  $\tau=-1$.
  
 The system thus looks like  an evolving  network, nodes being the GDP fluctuations or, in a short way, countries at different times. In Sec. III, we
 extract structures from the network through filtering the time delayed correlations (or distance matrix)  by removing the least correlated links. This percolation idea-based method reveals the emergence of connections, that are visualized by a branching representation. However the system is pretty unstable. Yet, expected features are reproduced. This should imply some further consideration in the line of  Glansdorff and  Prigogine \cite{Glansdorf}  about  Structure and Stability of world economic systems
and the intrinsic role of fluctuations.  Notice also that the network is much more complex when taking into account delay time. Indeed the correlation matrix is $NOT$ symmetrical,  whence the network links are not only weighted but also directed. The number of distances to be considered is $N(N-1)$, where $N$ is the number of countries, in contrast to $N(N-1)/2$, when $\tau=0$. In Sec. IV, we conclude that such a measure of collective habits does not fit the usual expectations defined by politicians but that does not imply that they are wrong or right, nor are we. 

\vspace{0.5cm}
We stress that several times must be truly considered, i.e. 

\begin{enumerate}
\item Initial observation time $t_0$
\item Data acquisition time $t$
\item Increment or fluctuation time span $\Delta $		 
\item Delay time (between two countries) $t^{'}$
\item Window observation (averaging) time $T $, i.e. [$t_{min}, t_{Max}$]
\item Conclusion time $t_N$
\end{enumerate}		
 
 Thus the correlation matrix is of quite high dimension ($i,j; t_0; t,t^{'},\tau$;...)

\section{Methodology}
We use data from $http://www.ggdc.net/index-series.html$
for
Normalized Gross Domestic Product (GDP) of the 21 most developed countries.  Most of countries have natural time scales associated to the time lags between elections. Whence we  use an average time of $T=4$ years. For each of the countries, let us define $p_{i;t}=\ln \frac{G_{i;t+T}}{G_{i;t}}$ that is equivalent to the average variation of the GDP over $T$ years. 

Let us now introduce a correlation measure and associated distance measure for the countries.
The correlation measure is based on the Theil index and is defined as follows
\begin{eqnarray}
C_{i,j;t,\tau} =\frac{<p_{i;t^{'}} ~ p_{j;t^{'}+\tau}>_t-<p_{i;t^{'}}>_t <p_{j;t^{'}+\tau}>_t}{\sqrt{<p_{i;t^{'}}^2 -<p_{i;t^{'}}>^2 >_t <p_{j;t^{'}+\tau}^2 -<p_{j;t^{'}+\tau}>^2 >_t}}
\end{eqnarray}
where the averages are defined as follows 
\begin{equation}
<F(t^{'})>_t = \frac{1}{\Delta} \sum_{t^{'}=t}^{t+\Delta} F(t^{'})
\end{equation}
The "averaging time" $\Delta$ has been hereby chosen to be $\Delta=4$ in the following.
The distance is obtained from the correlation from the definition:
\begin{equation}
d = \sqrt{\frac{1}{2} (1 - C)}
\end{equation}

Let us stress that this short hand writing  accounts for time delays between the signals. Let us also note that  other measures of distance between different countries can be defined, e.g

\begin{equation}
D_{i,j;t,\tau} = | p_{i;t} - p_{j;t + \tau}|
\end{equation},
but we will focus on the definition 1 and 3 in the present paper.
The comparison of results obtained for different measures will be considered elsewhere.

When $\tau=0$, the above quantities are at fixed time. When $\tau\neq0$, these quantities measure the correlations with a delay time between the economies of countries. 
If the evolutions were completely identical, one would find $D=0$.
Let us stress again that $D_{i,j;t,\tau}$ is symmetric in $i,j$, when $\tau=0$. For each pair of countries, this quantity defines therefore a surface whose statistical properties can be studied.

\begin{figure}
\includegraphics[width=5.0in]{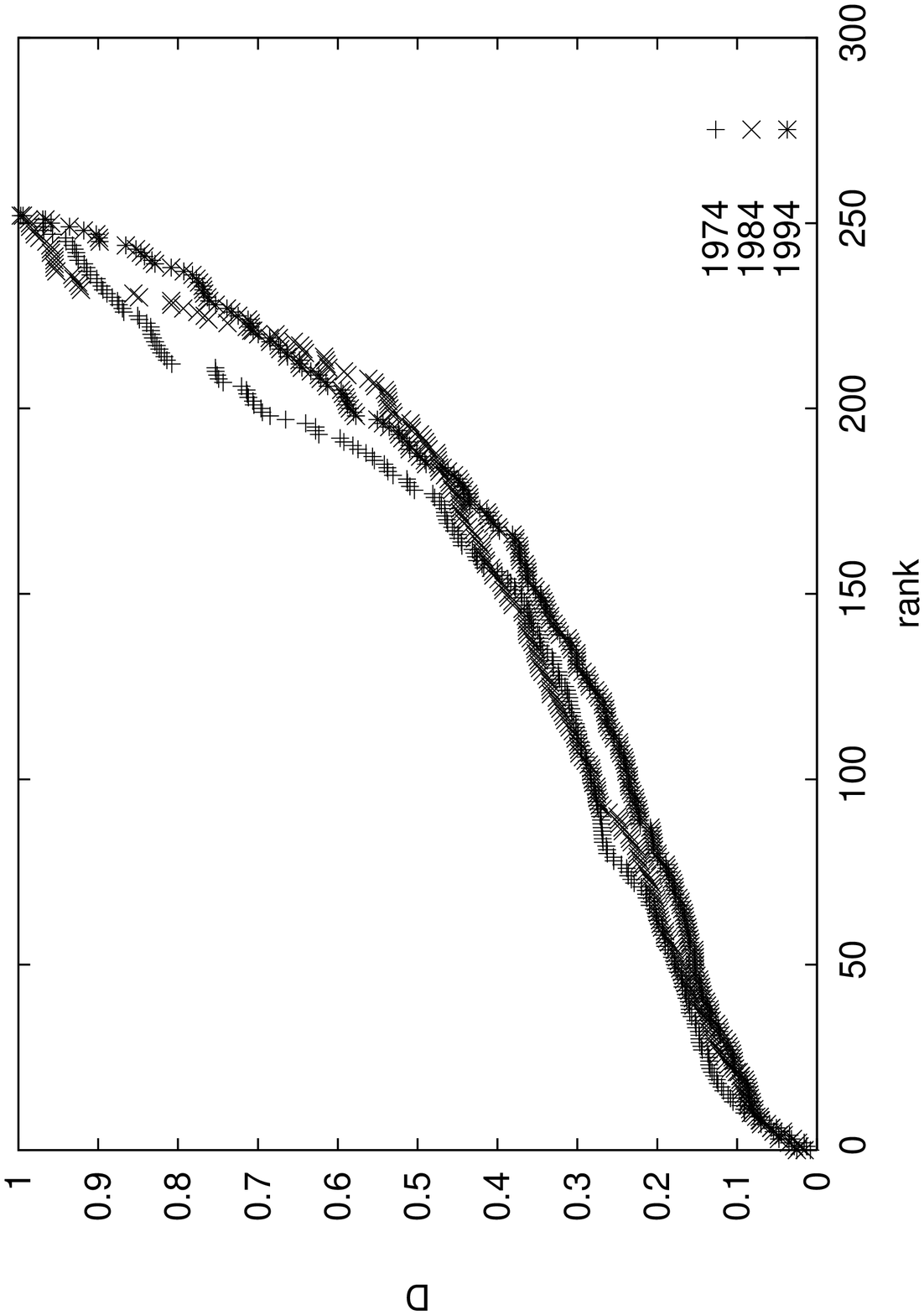}

\caption{\label{sim}  Rank analysis of $D_{i,j;t,\tau}$ for 1974, 1984, 1994 years when $\tau=0$}
\end{figure}

\begin{figure}
\includegraphics[width=5.0in]{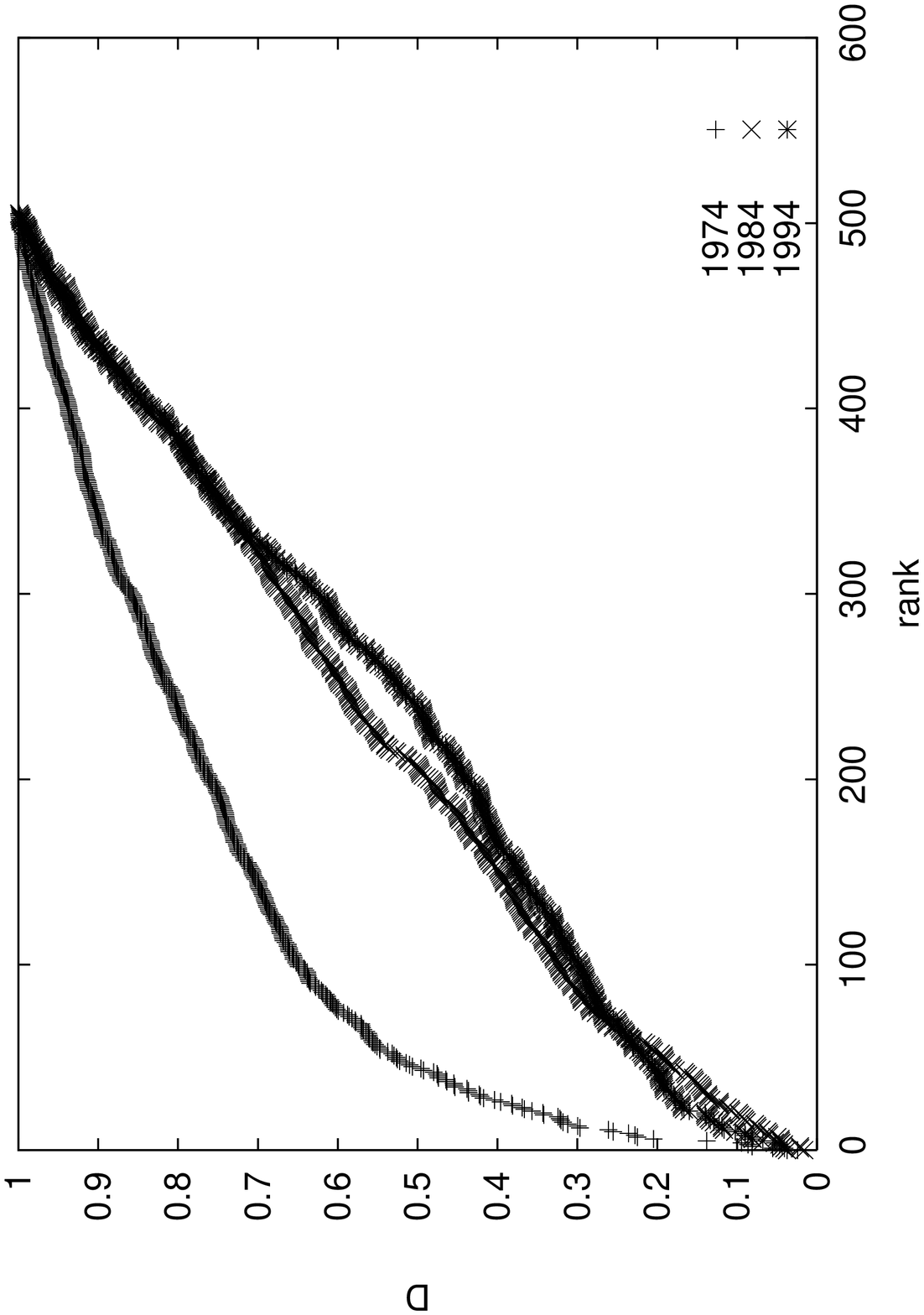}

\includegraphics[width=5.0in]{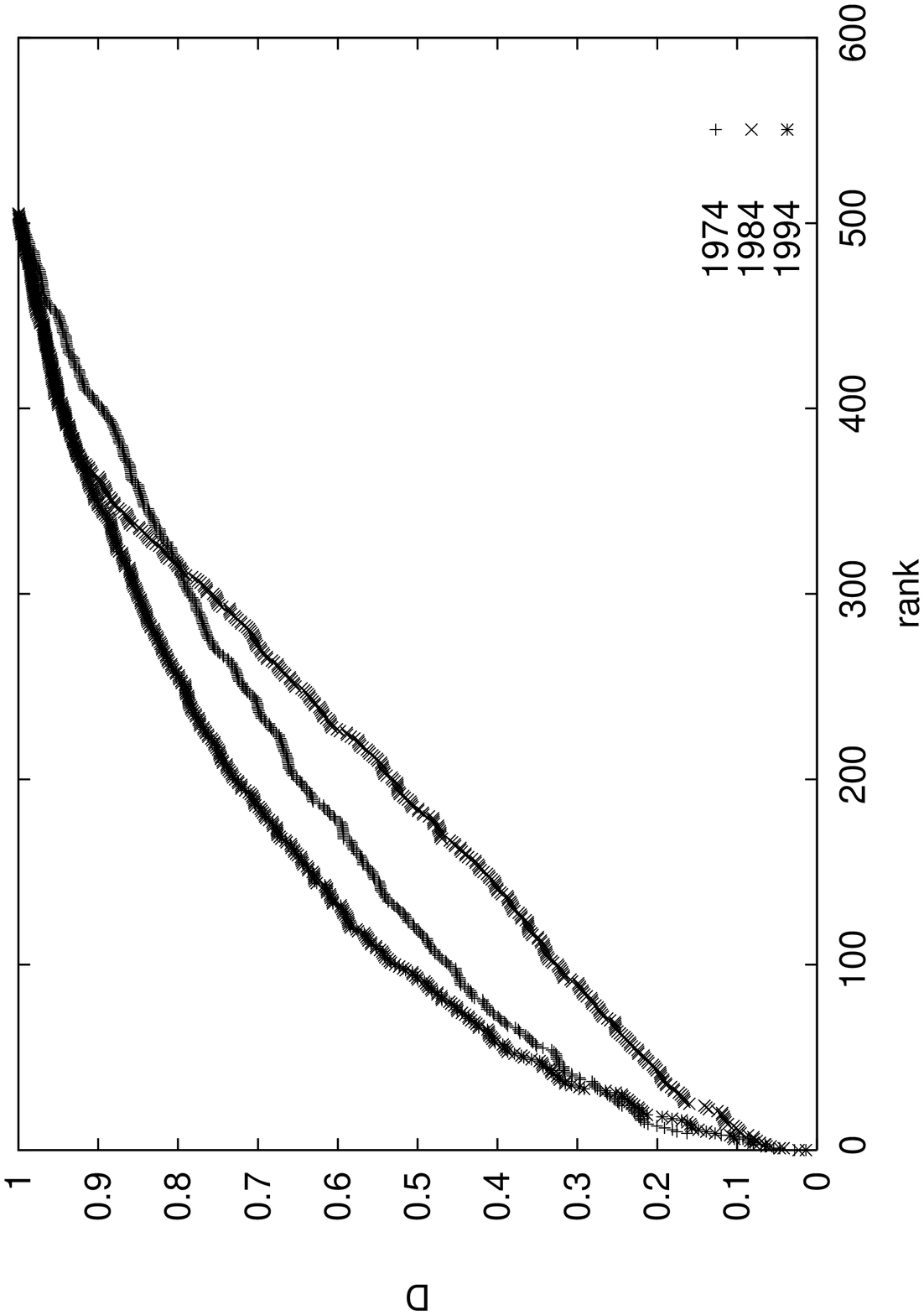}

\caption{\label{sim}  Rank analysis of $D_{i,j;t,\tau}$ for 1974, 1984, 1994  years,  when $\tau=1$ (upper figure) and when $\tau=-1$ (lower figure)}
\end{figure}

\begin{figure}
\includegraphics[width=2.5in]{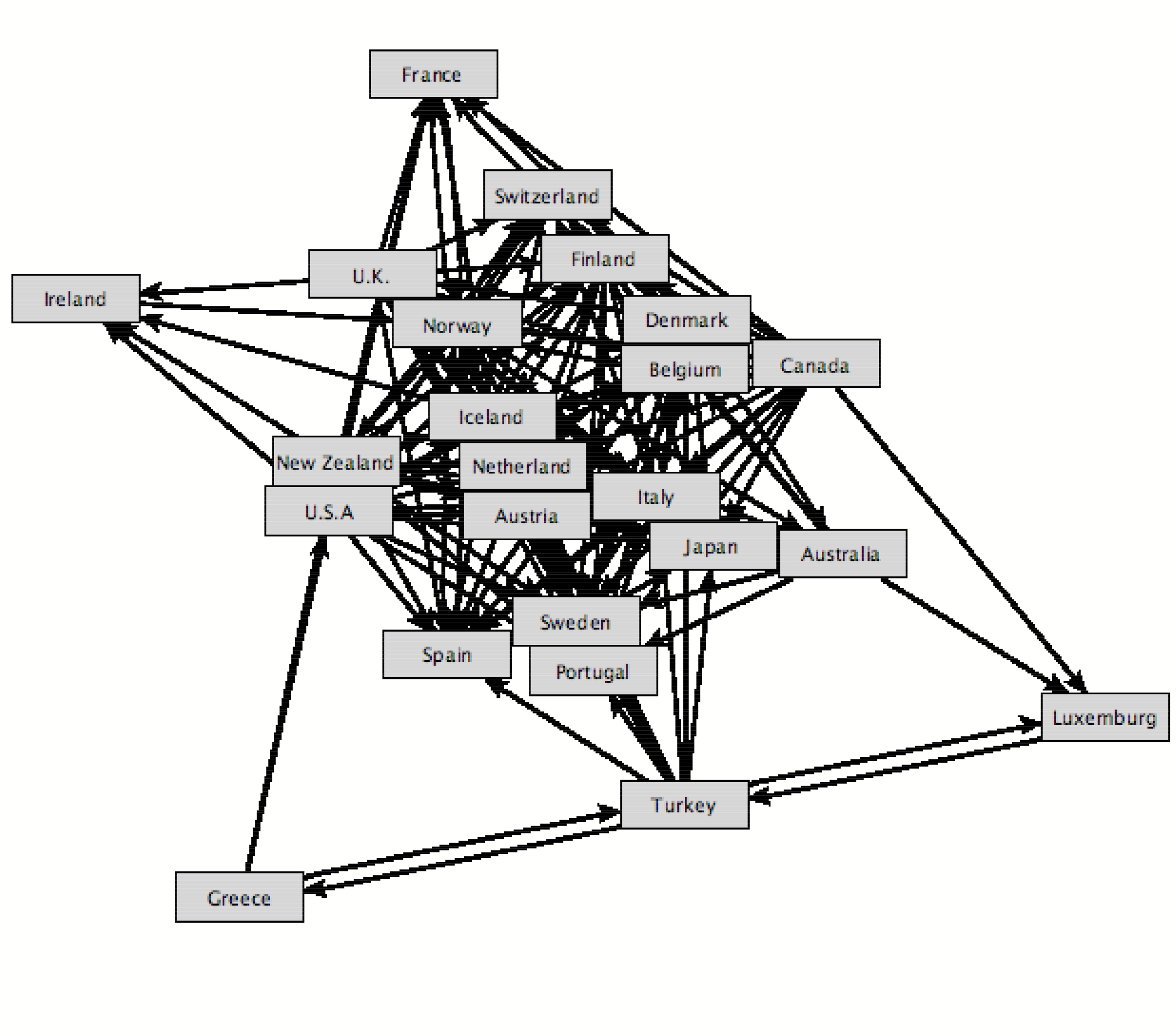}
\includegraphics[width=2.5in]{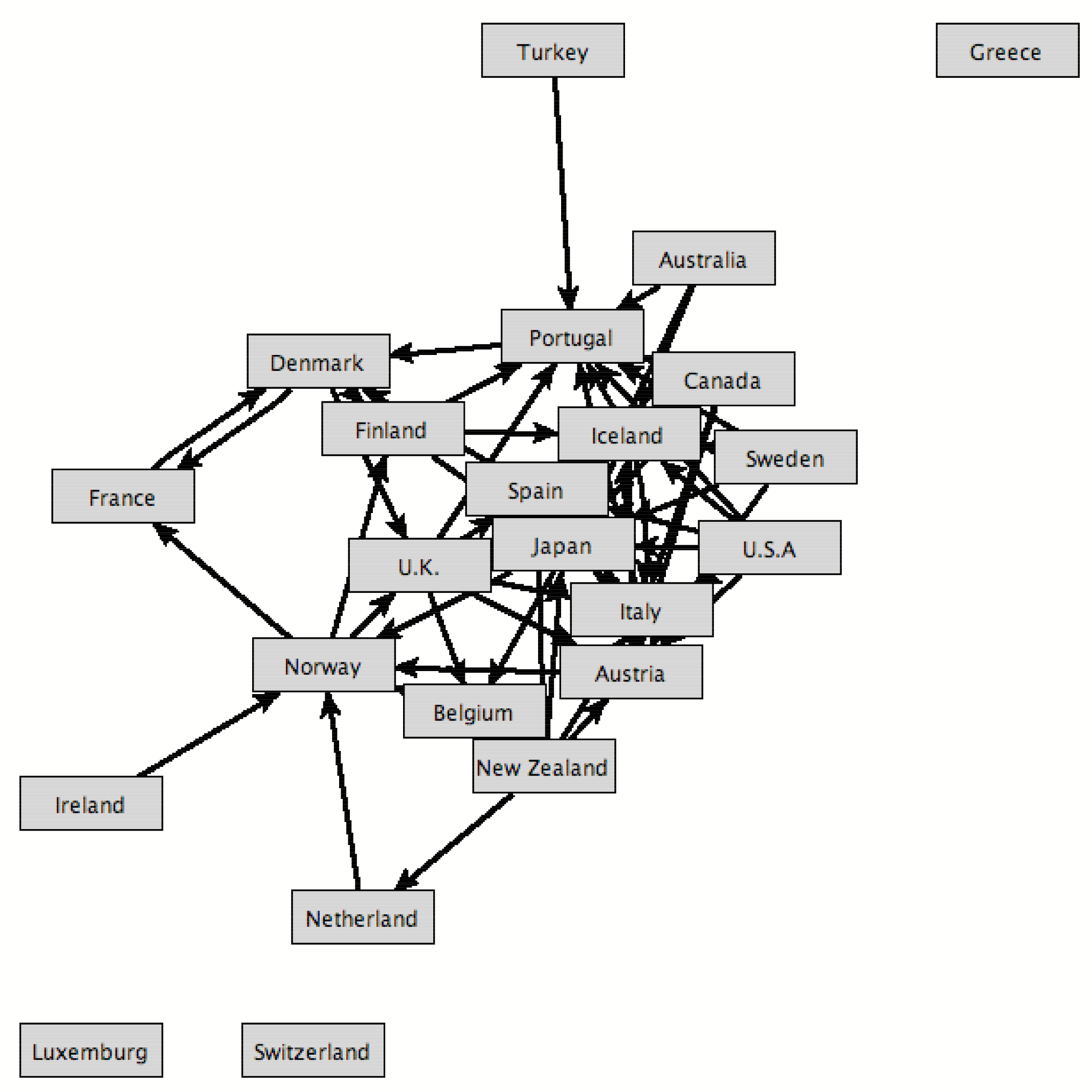}

\includegraphics[width=2.5in]{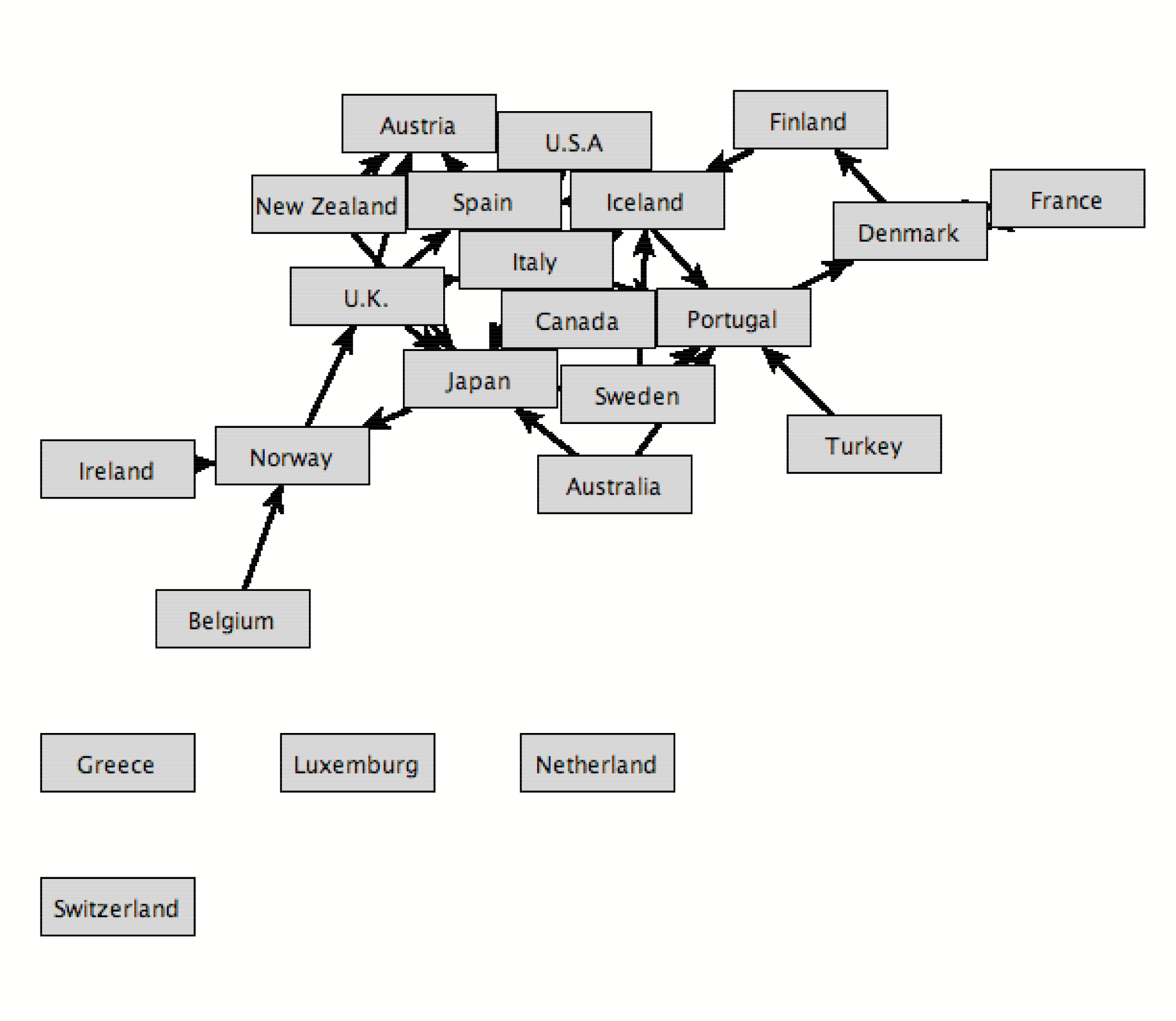}
\includegraphics[width=2.5in]{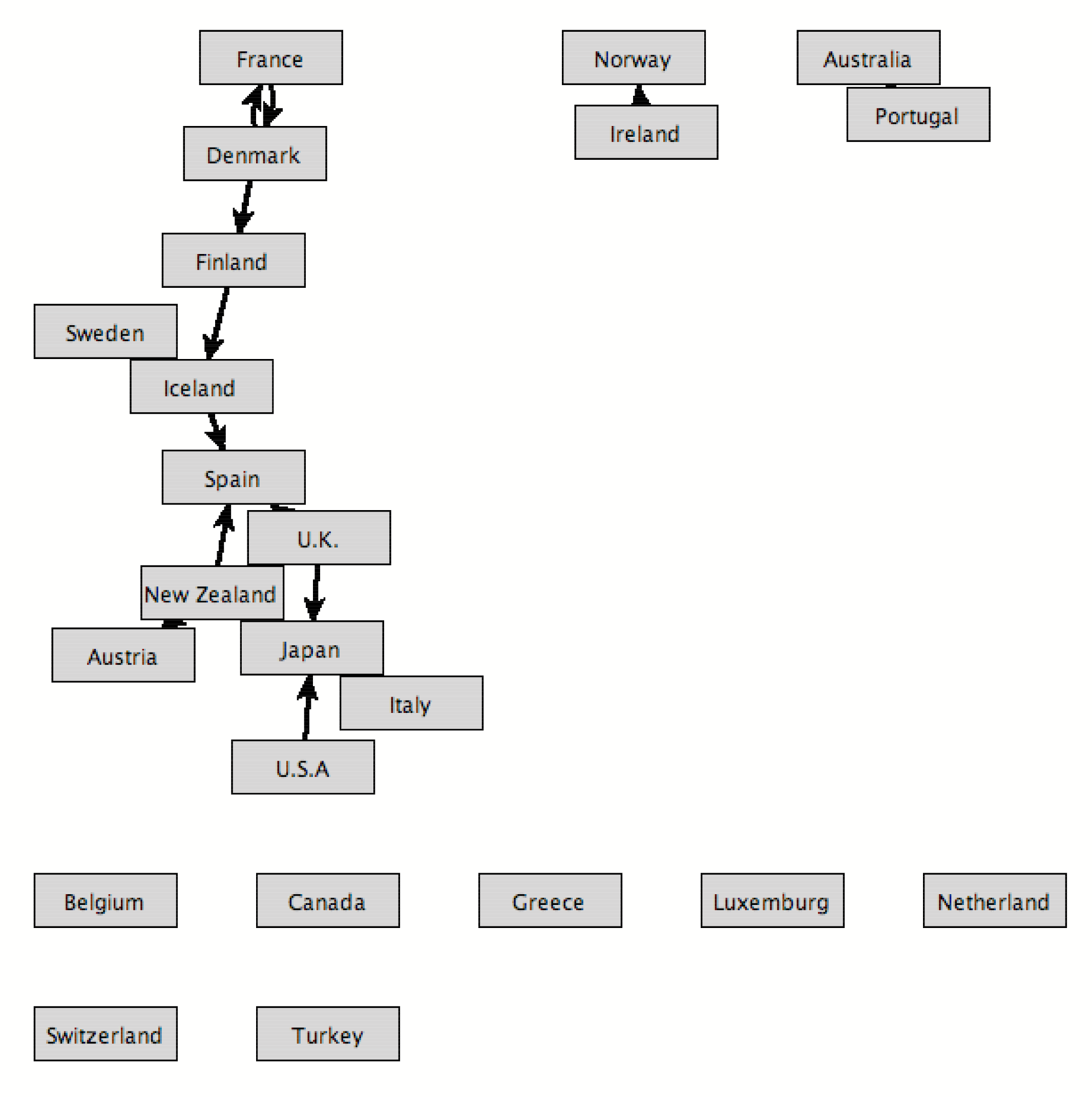}
\caption{\label{sim}  Clusters resulting from filtering  the distance matrix for $t=1984$ and $\tau=1$. The values of the filtering $f$ are respectively $0.45$, $0.20$, $0.15$ and $0.1$}
\end{figure}

The first relevant quantity to report is the rank analysis of  $D_{i,j;t,\tau}$, at different values of $t$ (1974, 1984, 1994) and $\tau$= 0, 1, -1.  (Figs. 1,2) 
Notice the non trivial behavior, with various crossing of the curves, and the  Zipf (log-log) plot, that does not lead to a fine straight line fit. Because the GDP increment distance between countries is rather uniform, this
suggests similarities in development patterns, which likely result from
interactions of economies, in a globalization sense. 

\section{Clustering analysis}

At fixed times, and fixed values of the delay $\tau$, pairs of countries (i,j) are characterized by their relative distance. In order to extract structures from the distance matrix, we build  country  networks through filtering the time delayed correlation matrix by removing the least correlated links through moving a threshold.

To do so, we define the filtered matrix $D^F$, where
\begin{eqnarray}
D^F_{ij}=0 , if ~~~ D_{ij} > f \cr
D^F_{ij}=1, if ~~~ D_{ij} < f
\end{eqnarray}
The resulting matrix therefore connects nodes that are very close to each other, and removes the weak links.
By decreasing  the values of $f$, one observes therefore  the breaking of the continent into several islands that correspond to very connected "communities of countries". Let us stress that this method has already been successfully applied in order to reveal structures in online communities \cite{lambi}. Contrary to this previous study,
 the distance matrix may be asymmetric when $\tau \neq 0$. Consequently, the network representation will be made of directed links from $i$ to $j$, if the matrix element $D_{ij} $ verifies $D_{ij} < f$. This percolation idea-based method reveals the emergence of connections, that are visualized by a branching representation. In Fig.4, this method is shown for the parameters $t=1984$, $\tau=1$.

\section{Conclusions}

 In this short paper, we have introduced techniques in order to study the development of structures in macro-economical systems. To do so, we have considered the GDP's of the 23 most developed countries, and defined statistical distances between their time evolution. A possible delay between economies has been accounted for through the parameter $\tau$, thereby allowing the distinction between {\em leading} and {\em following} countries.

In the above analysis, we have observed that the rank distribution is unexpectedly giving an exponent different from the classical Zipf law (=-1), and that the data contain a lot of noise. Nevertheless,
patterns have been found, under the form of clusters in the countries' network. This structuring seems to be in line with economy
globalization, that tends to homogenize the economic development of countries, and may be related to the influence of political considerations.

Clearly,  other values of the time lag should be examined, and not so rich
countries as well. There is no reason that the same time lag should be
considered for different countries. In fact one could look for higher
order time delayed correlation functions.

Finally, let us stress that there are many possible approaches, that could help the understanding of globalization processes in economy. A classical way in
searching for clusters is to find subgraphs with a high clustering coefficient
or to investigate the graph topology.

\section{Acknowledgments} This work is partially financially supported by ARC 02-07/293. RL would like also to thank  
  financial support from the European Commission Project 
CREEN FP6-2003-NEST-Path-012864. Thanks to J. Miskiewicz for fruitful comments.

\end{document}